# Quantum critical point and superconducting dome in the pressure phase diagram of *o-TaS₃*

M. Monteverde[1]*, J. Lorenzana[2], P. Monceau[1] and M. Núñez-Regueiro[1]

1 Institut Néel, Université Grenoble Alpes and Centre Nationale de la Recherche Scientifique, Grenoble, France

2 ISC-CNR, Dipartimento di Fisica, Sapienza, Università di Roma, Piazzale Aldo Moro 5, 00185 Roma, Italia

We measure the electrical resistance of $o-TaS_3$ between 1K and 300K under pressures up to 20GPa. We observe a gradual decrease of the charge density wave transition temperature $T_{CDW}$ with increasing pressure $P$ following a mean field quantum fluctuation power law $T_{CDW} = 215K\left[(P_c - P)/P_c\right]^\gamma$ with a quantum critical point at a pressure $P_c = 11.5 GPa$ and $\gamma \approx 0.5$. Around the quantum critical point we observe a superconducting dome with a maximum superconducting transition temperature $T_c = 3.1K$. Such dome is similar to superconducting domes around other types of order suggesting that the QCP is directly responsible for the enhancement of superconductivity through a universal mechanism still not well understood.

---

* Present address: Laboratoire de Physique des Solides, Université de Paris-Sud, Orsay, France

Linear transition metal chalcogenides have been thoroughly studied due to their one dimensional (1D) character. When metallic, the electronic bands yield almost flat Fermi (FS) surfaces that are prone to produce divergences in the Lindhard susceptibility at a wavevector $Q = 2k_F$ resulting in large electron-phonon coupling. Below a certain temperature, $T_{CDW} = 215K$, a charge density wave (CDW) develops, accompanied by a permanent distortion of the lattice at the same wavevector[1]. At the same time a BCS type gap develops on a large portion of the FS. Application of pressure enhances the hopping amplitude perpendicular to the chains, thus reducing the nesting[2], and hardens the phonon energy. As a result the CDW gets destabilized, gradually reducing its transition temperature. At some point, enough of the gapped FS occupied by the CDW is freed from its gap. In the simplest scenario the ungapped FS takes advantage of the electron-phonon coupling to develop a superconducting state[3] below a transition temperature $T_c$. Several linear chalcogenides, $NbSe_3$, $(TaSe_4)_2I$, $m-TaS_3$, $o-TaS_3$, have been studied and the existing measurements all follow the abovementioned description[4].

In recent years, interest has grown on what are called quantum phase transitions, i.e. phase transitions, from an ordered, e.g. ferromagnetic, to a disordered, e. g. paramagnetic, at $T = 0$ driven by an external parameter such as pressure. A huge amount of work has been done on heavy fermions, that normally have low ordering energy scales[5]. It has been found experimentally that these materials also show superconductivity when the ordering transition temperature is sufficiently depressed. Moreover, superconductivity very often develops below a dome centered at the quantum critical point (QCP), i.e. the point where the ordering temperature becomes zero. The ubiquity of superconducting domes around QCP's suggest that QCP's play an important role on promoting superconductivity. Otherwise it would not be clear why domes tend to be symmetrical around QCP with superconductivity weakening at both sides of the QCP. One view[6] is that the effective quasiparticle interaction becomes enhanced close to a QCP. A more recent proposal[7] is based on the idea that a QCP is a degenerate state of matter with a residual entropy which, quantum mechanics, in one way or another, will get rid of. Superconductivity appears as a natural candidate to quench the residual zero temperature entropy as it does not need of special features like nesting. Thus what makes QCP special may be not an enhanced pairing interaction but an enhanced pairing susceptibility close to a QCP[7]. The physics of QCP has been invoked to explain high-$T_c$ superconductivity in cuprates[6,8] even though a clear ordering under the superconducting dome has not yet been identified. One possibility is that the order is of the CDW type,[7] thus it is of great interest to find other examples where superconductivity is found around charge order to provide model systems where to study this interplay and challenge theories.

Whatever the correct explanation of the relation between QCP and superconductivity is, it is important to determine the exact dependence of $T_{CDW}$ with pressure, as the exponent of the observed power laws allows comparison with current theories[9]. Within this scope we have revisited the linear system $o-TaS_3$ in order to analyze in detail its pressure QCP, as previous work does not allow a precise determination of the power laws nor is unambiguous with respect to the coexistence of the CDW with superconductivity.

The electrical resistance measurements were performed using a Keithley 220 source and a Keithley 2182 nanovoltmeter. Pressure measurements, $1.4-22 GPa$ (between 4.2K and 300K), were done in a sintered diamond Bridgman anvil apparatus using a pyrophillite gasket and two steatite disks as the pressure medium[10]. Due to the solid state pressure medium, measurements can only be performed on compression. The $o-TaS_3$ samples are filamentary monocrystals from the same batch as those of Ref. 11, synthesized by H. Berger and loaned by F. Lévy, both from EPFL, Switzerland.

The structure of orthorhombic $o-TaS_3$ is still unknown. The lattice parameters are very large in the directions perpendicular to the chain direction c-axis, namely $a = 36.804$Å, $b = 15.173$Å, $c = 3.34$Å., and may indicate that it comprises 6x4 chains parallel to the c-axis. A single CDW transition develops [12] at $T_{CDW} = 215K$ with a CDW vector that is temperature dependent; its components[13] below $T_{CDW}$ are found [127] to be [$0.5a^*$, $(0.125 - \varepsilon)b^*$, $0.255c^*$].

We show on Fig. 1 the electrical resistivity as a function of temperature for different pressures. As expected, pressure destabilizes the CDW state, and the sample passes gradually from a semiconducting to a metallic ground state. Furthermore, above ~8GPa the sample shows superconductivity, whose onset increases up to a maximum of ~3.3K at the pressure where the CDW disappears altogether (Fig. 2).

We have measured the critical magnetic field of the superconducting state at different temperatures. The critical magnetic field at zero temperature $H_C(0,P)$ can then be calculated at each pressure by fitting the expression:

$$H_C(T,P) = H_C(0,P)\left(1 - \left(\frac{T}{T_C(P)}\right)^2\right) \quad [1]$$

The pressure dependence of the correlation length can be obtained from the Ginzburg-Landau expression:

$$\xi(P) = \left(\frac{\phi_0}{2\pi H_C(0,P)}\right)^{1/2} \quad [2]$$

where $\phi_0$ is the magnetic flux quantum. No significant variation is observed on the correlation length as function of pressure (Fig. 2).

The transition temperature to the CDW state, $T_{CDW}$, can be determined by the logarithmic derivative $d\log R/dT$. However, we have observed that the same derivative but with respect to the inverse temperature, $d\log(R)/d(1/T)$, is a more sensitive method that allows to follow $T_{CDW}$ down to stronger pressures, while coinciding with the other definition at lower pressures. We have adopted thus this method (Fig. 3).

In Fig. 4 we show the measured pressure phase diagram of $o-TaS_3$. We compare our results to older measurements with a good agreement between them. However, we are now able to determine precisely the pressure dependence of $T_{CDW}$. Astonishingly, in all the pressure range, it follows a mean field quantum fluctuation power law $T_{CDW} = 215K\left[(P_c - P)/P_c\right]^\gamma$ with a critical pressure $P_c = 11.5 GPa$ and $\gamma = 0.5$. This is unexpected, as for $NbSe_3$, a compound of the same 1D transition metal trichalcogenide family but with a simpler structure, according to the reported results[14,15,16], pressure yields first an exponential decrease that, only when $T_{CDW}$ becomes small enough, changes towards a $\gamma = 0.5$ power law. The exponential decrease is expected due to the BCS dependence[1] of $T_{CDW} = 5.43 T_F \exp[-\omega_q / g^2 N(E_F)]$, where $T_F$ is the Fermi energy, $\omega_q$ the frequency of the phonon that stabilizes the CDW, $g$ the coupling parameter and $N(E_F)$ the density of states at the Fermi level. The most important variation will be given by the exponential, where $\omega_q$ will obviously increase with pressure, while $N(E_F)$ is expected to decrease at the same time as should $g$. Supposing a linear variation of these parameters, we would expect an exponential decrease with pressure of $T_{CDW}$ until we are sufficiently near to the QCP, where fluctuations take over in the form of a power law $T_{CDW} \propto \left[(P_c - P)/P_c\right]^{0.5}$. This is the type of behavior observed in $NbSe_3$ and that has been also studied in detail in Cr metal[17], whose known itinerant antiferromagnetism is due to a BCS type of spin density wave. According to common belief the expectation for the dependence of $T_{CDW}$ with pressure is: an exponential decrease with pressure characteristic of a weak coupling BCS state, followed, at sufficiently low transition temperatures by a power low decrease due to quantum fluctuations. $o-TaS_3$ seems thus to be anomalous, without regime of exponential pressure dependence, in spite of a high CDW transition temperature. Apparently quantum fluctuations control in same unexpected way the variation with pressure of $T_{CDW}$ far away from the QCP.

Another important point of our measurements is the superconducting dome that we observe around the QCP within our error bars. In the field of heavy fermions, a large amount of discussion has been developed around the question if superconductivity is due to the quantum critical point. The carriers would forestall the intense critical fluctuations of the quantum critical point by re-organizing themselves into a new stable phase of matter, i.e. superconductivity [18,19]. However, in the CDW case, it is natural to expect the appearance of superconductivity as the CDW disappears, as given by the Bilbro-McMillan[3] formula $T_{CDW}^{1-n} T_c^n = T_{c0}$, where $n$ is the superconducting fraction of the Fermi surface and $T_{c0}$ the superconducting transition temperature without CDW. However, this formula, that does not predict a superconducting dome, is obtained considering that BCS controls the evolution of both transition temperatures. We have seen that for $o-TaS_3$, in the entire the phase diagram, and for , $NbSe_3$ near the QCP, fluctuations control $T_{CDW}$. Thus it is clear that new theoretical ideas which could lead to a superconducting dome[6] are most welcome as well as further theoretical and experimental studies which could assert their applicability to CDW systems.

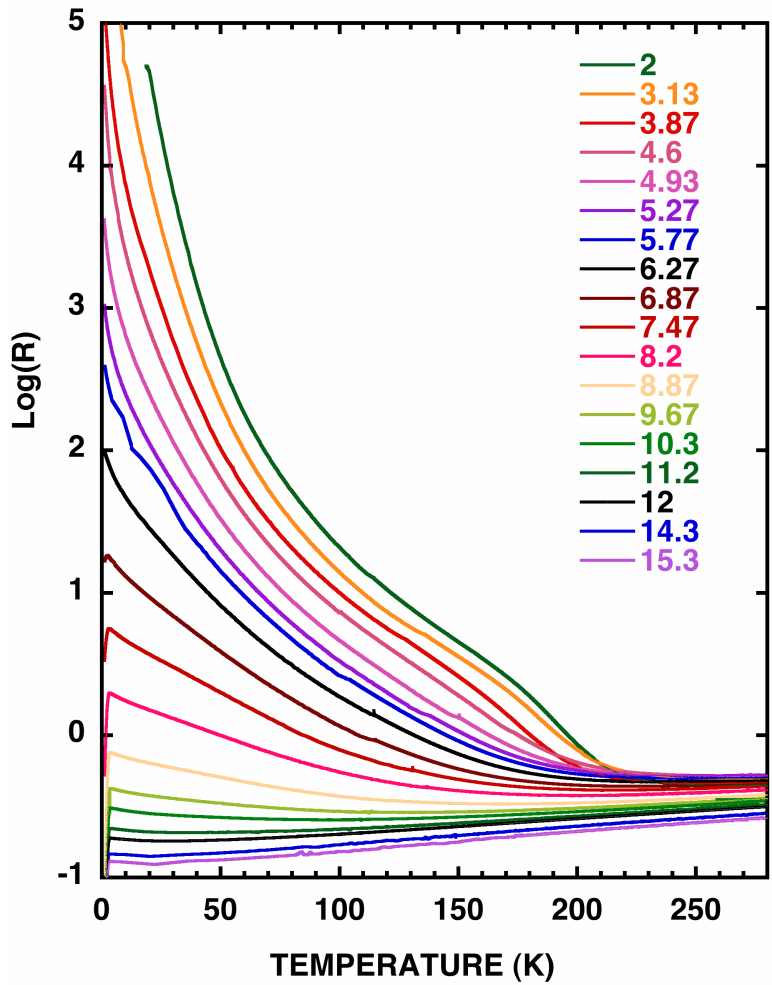

**Figure 1**

Logarithm of the electrical resistivity of $o-TaS_3$ as a function of temperature for different pressures. The passage from a semiconducting low temperature ground state towards a metallic and superconducting state is apparent.

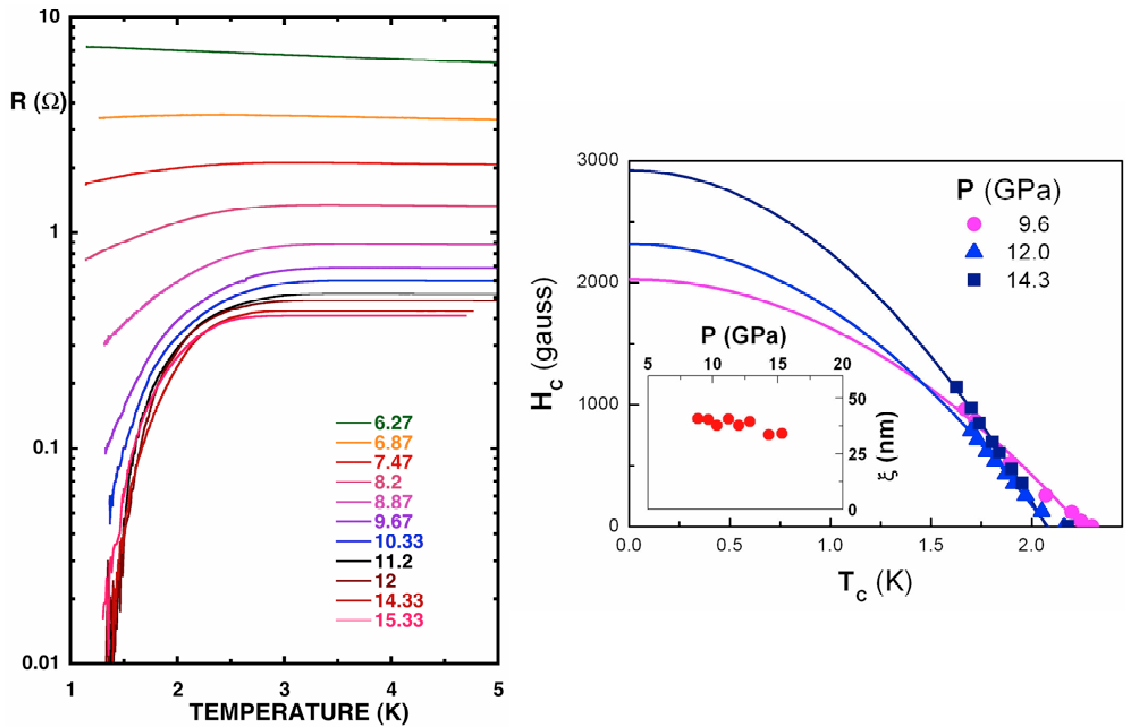

**Figure 2**

Left Panel: Low temperature zoom of the logarithm of the electrical resisitivity of $o-TaS_3$ for different pressures, showing the transitions to a superconducting state. Right panel: Dependence with temperature of the Critical magnetic field (for clarity only three pressures are shown). The continuous curves are fit to the data according to expression (1). Insert: Correlation length as function of pressure obtained from the Critical magnetic field of each pressure.

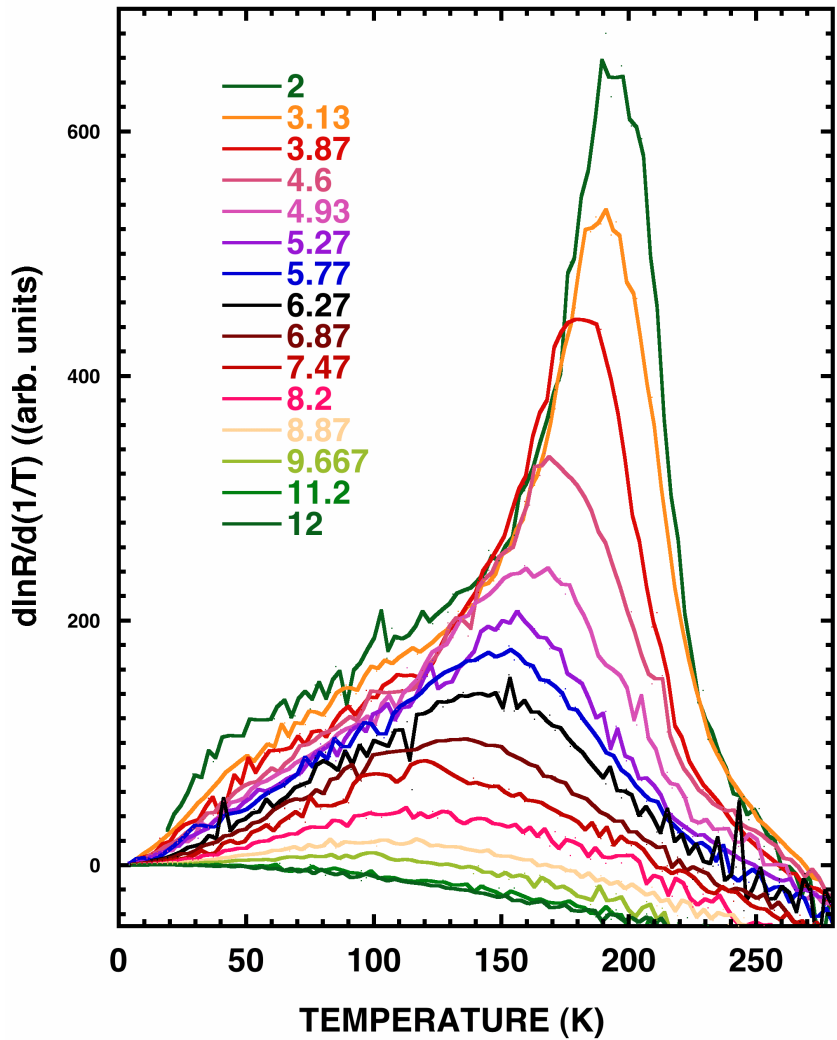

**Figure 3** The logarithmic derivative of the electrical resistivity with respect of the inverse temperature, $d\log(R)/d(1/T)$ for different pressures. This is the most sensitive method to follow the evolution of the CDW transition with pressure. It is defined by the temperature that corresponds to the maximum in each curve.

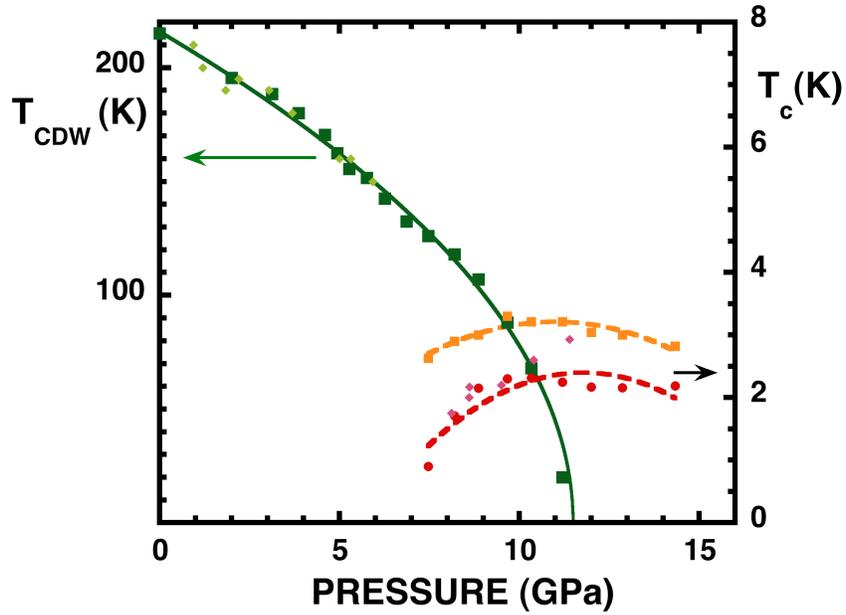

**Figure 4** Pressure phase diagram of $o-TaS_3$. $T_{CDW}$'s ($T_c$'s) obtained in this work are shown as green squares (orange squares for the onset and red circles for 50% of the superconducting transition) together to previous reports[4] (diamonds). We see that there is a good agreement between both measurements. $T_{CDW}$ follows a mean field power law $T_{CDW} \propto \left[(P_c - P)/P_c\right]^{0.5}$. Within our limited low temperature range, superconductivity appears within a dome surrounding the QCP, and coexists with the CDW below $P_c$.